\documentstyle[12pt,epsfig]{article}

\newcommand{\alr}{A_{LR}}
\newcommand{\alro}{A_{LR}^0}
\newcommand{\swein}{\sin^2\theta_W^{\rm eff}}
\newcommand{\polt}{{\cal P}_\tau}
\newcommand{\pole}{{\cal P}_e}
\newcommand{\polp}{{\cal P}_p}
\newcommand{\poll}{{\cal P}}
\newcommand{\eff}{\varepsilon}
\newcommand{\updates}[1]%
 {\fbox{\parbox{\linewidth}{\textbf{Updates with respect to last year:}\\#1}}}
\def\pz{\phantom{0}}

\def\ifmath#1{\relax\ifmmode #1\else $#1$\fi}%
\def\GeV{\ifmmode {\mathrm{ Ge\kern -0.1em V}}\else
                   \textrm{Ge\kern -0.1em V}\fi}%
\def\MeV{\ifmmode {\mathrm{ Me\kern -0.1em V}}\else
                   \textrm{Me\kern -0.1em V}\fi}%
\def\keV{\ifmmode {\mathrm{ ke\kern -0.1em V}}\else
                   \textrm{ke\kern -0.1em V}\fi}%
\def\eV{\ifmmode  {\mathrm{ e\kern -0.1em V}}\else
                   \textrm{e\kern -0.1em V}\fi}%

\def\mca#1#2 {\multicolumn{#1}{|c|}{#2}}

\newcommand{\nb}{\rm{nb}}

\newcommand{\Afbpol}{A^{0,\,\ell}_{\rm {FB}}}

\newcommand{\Afbze}{A^{0,\,{\rm e}}_{\rm {FB}}}

\newcommand{\Afbzm}{A^{0,\,\mu}_{\rm {FB}}}
\newcommand{\Afbzt}{A^{0,\,\tau}_{\rm {FB}}}

\renewcommand{\rm}{\mathrm}

\newcommand{\MZ}{m_{\mathrm{Z}}}
\newcommand{\MW}{m_{\mathrm{W}}}

\newcommand{\MH}{m_{\mathrm{H}}}
\newcommand{\Mt}{m_{\mathrm{t}}}

\newcommand{\GZ}{\Gamma_{\mathrm{Z}}}

\newcommand{\RZ}{R_{\ell}}
\newcommand{\Ree}{R_{\mathrm{e}}}
\newcommand{\Rmu}{R_{\mu}}
\newcommand{\Rtau}{R_{\tau}}

\newcommand{\ee}{\mathrm{e}^+\mathrm{e}^-}

\newcommand{\Gee}{\Gamma_{\rm {ee}}}

\newcommand{\Gmumu}{\Gamma_{\mu\mu}}
\newcommand{\Gtautau}{\Gamma_{\tau\tau}}
\newcommand{\Ginv}{\Gamma_{\mathrm{inv}}}
\newcommand{\Ghad}{\Gamma_{\mathrm{had}}}

\newcommand{\Gll}{\Gamma_{\ell\ell}}

\newcommand{\shad}{\sigma_{\mathrm{h}}^{0}}

\newcommand{\cAe}{\mbox{$A_{\rm e}$}}
\newcommand{\cAt}{\mbox{$A_{\tau}$}}

\def\Ups4s{\mbox{$\Upsilon(4S)$}}
\newcommand{\etal}{\mbox{{\it et al.}}}
\def\Title#1{\begin{center} {\Large {\bf #1} } \end{center}}

\begin{document}

\title{Precision Electroweak Physics at the $Z$} 
\author{\\ Morris L. Swartz\\ \\
{\it Department of Physics and Astronomy}\\
{\it The Johns Hopkins University}\\
{\it 3400 North Charles Street}\\
{\it Baltimore, Maryland 21218 U.S.A.}}
\maketitle
\thispagestyle{empty}
\setcounter{page}{0}
\begin{abstract}
A review of the decade of $Z$-pole electroweak physics is presented.  Although all experimental work has been completed, it represents a ``Golden Age'' in our understanding of the Minimal Electroweak Standard Model (MSM).  The latest (and nearly final) results from the LEP and SLC experiments are presented.  The remaining inconsistencies are discussed and and their effects upon the MSM interpretation are explored. 
\end{abstract}
\bigskip\bigskip\bigskip
\centerline{Plenary talk presented at the}{ XIX International
Symposium on Lepton and Photon Interactions at High Energies}
\centerline{Stanford University, August 9-14, 1999}
\vfill\eject

\Title{Precision Electroweak Physics at the $Z$}

\bigskip\bigskip


\begin{raggedright}  

{\it M.L. Swartz\index{Swartz, M.L.}\\
Department of Physics and Astronomy \\
Johns Hopkins University,
Baltimore, Maryland 21218 }
\bigskip\bigskip
\end{raggedright}

\section{Introduction}

The experimental study of the process $\ee\to Z\to f\bar f$ began 10 years ago and ended (data taking) just over 1 year ago.  The first electroweak measurements performed at the $Z$ were presented exactly 10 years ago at LP89 in this auditorium \cite{ref:markii}.  Those results which were based upon a sample of 233 events are compared with the present state of the art in Table~\ref{tab:lpcomparison}.  
The LP99 measurements are approximately 100 times more precise than their LP89 counterparts!  The reader should keep in mind that the LP89 measurements were among the best electroweak measurements available at that time.  
\begin{table}[htb]
\begin{center}
\begin{tabular}{|lcc|}
\hline
Quantity & LP89 (233 events) & LP99 (18M events) \\
\hline
$\MZ$ (GeV) & 91.17$\pm$0.18 & 91.1871$\pm$0.0021 \\
$\GZ$ (GeV) & 1.95$^{+0.40}_{-0.30}$ & 2.4944$\pm$0.0024 \\
$N_\nu$ & 3.0$\pm$0.9 & 2.9835$\pm$0.0083 \\ \hline
\end{tabular}
\caption{A comparison the $Z$-pole electroweak results presented at LP89 and the similar results at LP99.}
\label{tab:lpcomparison}
\end{center}
\end{table}

The last decade has seen a remarkable improvement in our knowledge of various electroweak parameters.  Much of the improvement is due to the study of the $Z$ resonance at LEP and the SLC.  The LEP program completed data-taking at the $Z$ in 1995 and the SLC program finished in 1998.  Many final or nearly final electroweak results are now being produced by the five experiments: ALEPH, DELPHI, L3, OPAL, and SLD.  These new results are the subject of this talk which may well be the last Lepton-Photon talk dedicated to $Z$ pole electroweak physics.

This document is organized as follows: Section~\ref{sec:defs} includes definitions of the quantities to be discussed, Section~\ref{sec:results} contains a summary of the experimental results, Section~\ref{sec:interpretation} contains a brief interpretation of the results, and Section~\ref{sec:summary} contains a summary and conclusions. 

\section{Definitions}
\label{sec:defs}

\subsection{The $Zf\bar f$ Vertex}

The coupling of the $Z$ to a fermion ($f$) antifermion ($\bar f$) pair is described by the following Lagrangian density,

\begin{eqnarray}
{\cal L} &=& \left(\frac{G_F\MZ^2}{2\sqrt{2}}\right)^{1/2}\overline{\Psi}_f\gamma_\mu\left(v_f-a_f\gamma_5\right)\Psi_fZ^\mu \label{eq:lagrang}\\
&=&
\left(\frac{G_F\MZ^2}{2\sqrt{2}}\right)^{1/2}\overline{\Psi}_f\gamma_\mu\left[g_L^f\left(1-\gamma_5\right)+g_R^f\left(1+\gamma_5\right)\right]\Psi_fZ^\mu 
\end{eqnarray}
where $v_f$ and $a_f$ are vector and axial vector coupling constants, and $g_L^f=(v_f+a_f)/2$ and $g_R^f=(v_f-a_f)/2$ are left- and right-handed combinations.  The vector and axial vector couplings are related to the quantum numbers of the fermion as follows
\begin{eqnarray}
v_f &=& \sqrt{\rho_f}\left(2I_3^f-4Q_f\sin^2\theta_f\right)  \\
a_f &=& \sqrt{\rho_f}\left(2I_3^f\right).
\end{eqnarray}
where $I_3^f$ is the third component of weak isospin, $Q_f$ is the electric charge, and the parameters $\rho_f\sim1$ and $\sin^2\theta_f\sim0.23$ incorporate electroweak radiative corrections.

\subsection{The $Z$-Peak Cross Section}

The cross section for the process $e^+e^-(\pole)\to Z\to f\bar f$ is described in the center-of-mass frame by the following expression,
\begin{equation}
\frac{d\sigma_Z^f}{d\Omega} = \frac{9}{4} \frac{s\Gamma_{ee}\Gamma_{f\bar f}/\MZ^2}{(s-\MZ^2)^2+s^2\GZ^2/\MZ^2} \left\{ 
 \left(1+\cos^2\theta\right)\left[1-\pole A_e\right] + 2\cos\theta A_f\left[-\pole+A_e\right] \right\} \label{eq:zpolexs}
\end{equation}
where: $\pole$ is the polarization of the electron beam, $s$ is the square of the cm energy, $\MZ$ is the mass of the $Z$, $\GZ$ is the total width of the $Z$, $\theta$ is the angle between the incident electron and the outgoing fermion, $\Gamma_{f\bar f}$ is the partial width for $Z\to f\bar f$, and $A_f$ is the left-right coupling constant asymmetry.  The partial widths and coupling coupling constant asymmetries are related to the couplings defined in the Lagrangian,
\begin{eqnarray}
\Gamma_{f\bar f} &=& N_c\cdot\frac{G_F\MZ^3}{24\pi\sqrt{2}}\left(v_f^2+a_f^2\right)\left(1+\delta_{rc}\right)  \\
A_f &=& \frac{2v_fa_f}{v_f^2+a_f^2} = \frac{(g_L^f)^2-(g_R^f)^2}{(g_L^f)^2+(g_R^f)^2}, 
\end{eqnarray}
where $\delta_{rc}\simeq1+3\alpha Q_f^2/4\pi+\eta_f\alpha_s/\pi$ ($\eta_f$ is 1 for quarks and 0 for leptons) accounts for final state radiative effects.

The small size of $v_\ell/a_\ell$ ($\sim0.08$) makes the leptonic coupling asymmetries $A_\ell$ particularly sensitive to electroweak vacuum polarization corrections.  The leptonic asymmetries are usually parameterized in terms of $\swein=\sin^2\theta_\ell$ (assuming lepton universality).  It follows that small changes in $\swein$ produce large effects on $A_\ell$,
\begin{eqnarray}
A_\ell &=& \frac{2(1-4\swein)}{1+(1-4\swein)^2} \\
\delta A_\ell &\simeq& -8\delta\swein
\end{eqnarray}

\subsection{Z-Pole Electroweak Observables}

\begin{figure}[htb]
\begin{center}
\epsfig{file=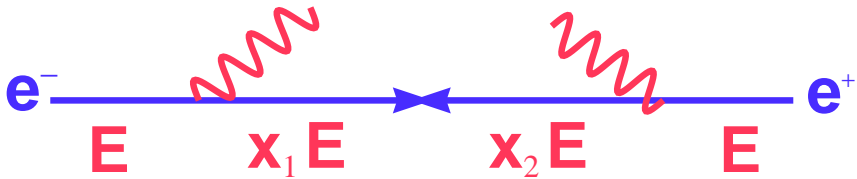}
\caption{The effect of initial state radiation on the center of-mass energy of the $f\bar f$ system.}
\label{fig:inirad}
\end{center}
\end{figure}

The cross section described in equation~\ref{eq:zpolexs} is only the dominant term in the total s-channel $e^+e^-$ cross section which can be expressed as
\begin{equation}
\frac{d\sigma_{tot}^f}{d\Omega}(s) = \frac{d\sigma_Z^f}{d\Omega}(s)+\frac{d\sigma_{Z\gamma}^f}{d\Omega}(s)
+\frac{d\sigma_\gamma^f}{d\Omega}(s),
\end{equation}
where the second and third terms represent $Z\gamma$ interference and pure $\gamma$ exchange, respectively.  The presence of initial-state radiation smears the center-of-mass energy (see Figure~\ref{fig:inirad}) so that the observed cross section can be represented by the integral
\begin{equation}
\frac{d\sigma_{obs}^f}{d\Omega}(s)=\int dx_1 dx_2 D_e(x_1,s) D_e(x_2,s)\frac{d\sigma_{tot}^f}{d\Omega}(sx_1x_2) \label{eq:xsobs}
\end{equation}
where the electron structure function $D(x,s)$ represents the probability that the incident electron (or positron) radiates the fraction $1-x$ of its energy before interacting with the other particle.  

Using the $Z$-pole cross section given in equation~\ref{eq:zpolexs} it is possible to define a number of experimental observables (or pseudo-observables since the $Z$-pole cross section isn't exactly the observed cross section):
\begin{enumerate}
\item
the line shape parameters which consist of the $Z$ mass $\MZ$, the total width $\GZ$, and the peak hadronic cross section $\shad=12\pi\Gamma_{ee}\Gamma_{had}/(\MZ^2\GZ^2)$;
\item
the cross section ratios $R_\ell = \Gamma_{had}/\Gamma_\ell$, $R_b = \Gamma_{bb}/\Gamma_{had}$, and $R_c = \Gamma_{cc}/\Gamma_{had}$;
\item
the unpolarized forward-backward asymmetries which are defined as
\begin{equation}
A_{FB}^f = \frac{\sigma_F^f-\sigma_B^f}{\sigma_F^f+\sigma_B^f} = 0.75A_eA_f,~~~f=\ell,b,c
\end{equation}
where $\sigma_F^f$ is the cross section for finding the scattered fermion in the hemisphere defined by the incident electron direction and $\sigma_B^f$ is the cross section for finding it in the positron hemisphere;
\item
the left-right asymmetry which is defined as
\begin{equation}
A_{LR}^m = \frac{\sigma^f(-|\pole|)-\sigma^f(+|\pole|)}{\sigma^f(-|\pole|)+\sigma^f(+|\pole|)} = \pole\alro = \pole A_e,~~~f\neq e 
\end{equation}
where $\sigma^f(\pole)$ is total (angle integrated) cross section for the production of $f\bar f$ pairs with an electron beam of helicity $\pole$;
\item
the polarization of final state $\tau$-leptons which depends upon the direction of the $\tau$, $A_e$, $A_\tau$,
\begin{equation}
{\cal P}_\tau(\cos\theta) = -\frac{A_\tau(1+\cos^2\theta)+2A_e\cos\theta}{1+\cos^2\theta+2A_\tau A_e\cos\theta};  \label{eq:taupol}
\end{equation}
\item
and the left-right forward-backward asymmetries which are defined as
\begin{eqnarray}
\tilde A_{FB}^f &=& \frac{\sigma_F^f(-|P|)-\sigma_B^f(-|P|)-\sigma_F^f(+|P|)+\sigma_B^f(+|P|)} {\sigma_F^f(-|P|)+\sigma_B^f(-|P|)+\sigma_F^f(+|P|)+\sigma_B^f(+|P|)} \nonumber \\ \nonumber \\ &=& 0.75PA_f,~~~f=\ell,b,c,s.
\end{eqnarray}
\end{enumerate}

The observed cross section near $s=\MZ^2$ does not differ dramatically from the resonance cross section given in equation~\ref{eq:zpolexs} because: the $Z$-$\gamma$ interference cross section vanishes at the pole, the $\gamma$-exchange cross section is approximately 1000 times smaller than the $Z$-exchange cross section, and the electron structure functions are strongly peaked near $x=1$.  The net correction to measured observables varies from less than 2\% for $\tau$-polarization and the left-right asymmetry to about 30\% for the $Z$-peak cross section to about 100\% for the leptonic forward-backward asymmetries.  These corrections are usually calculated from the Minimal Standard  Model (MSM) with the assumption that the MSM is an adequate description of the (dominant) interference and pure $\gamma$-exchange corrections.

\section{Results}
\label{sec:results}

\subsection{The $Z$ Resonance Parameters}
\label{sec:zlineshape}

The LEP experiments measure the resonance parameters $\MZ$, $\GZ$, $\shad$, $R_\ell$ with final state hadronic and leptonic samples collected during scans of the $Z$ peak.  Since the leptonic forward-backward asymmetries are sensitive functions of $\sqrt{s}$, the $A_{FB}^\ell$ are also extracted from a simultaneous fit to hadronic and leptonic lineshape data. 

\begin{figure}[hbt]
\begin{center}
\epsfig{file=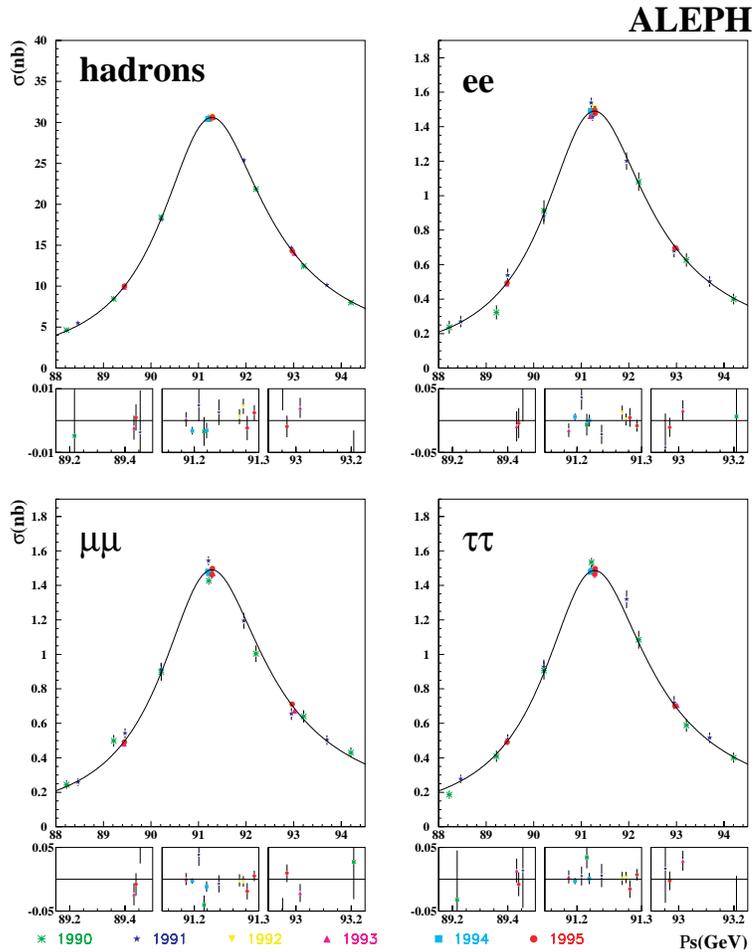,width=4in}
\caption{The hadronic and leptonic lineshape data of the ALEPH Collaboration.}
\label{fig:alephline}
\end{center}
\end{figure}

All four LEP experiments have recently updated their lineshape results \cite{ref:lepew}.
The total LEP event sample consists of $15.5\times10^6$ $Z\to q\bar q$ and $1.7\times10^6$ $Z\to\ell^+\ell^-$ events collected at $\sim$7 energies from 1990 to 1995.
The $Z\gamma$ and pure $\gamma$ cross sections are fixed to MSM~values and the data fit to ($3^{rd}$-order) radiatively-corrected lineshape functions.  The ALEPH lineshape fits are shown in Fig.~\ref{fig:alephline} \cite{Barate:1999ce}.

The leptonic parameters are determined separately for each lepton species (9 lineshape parameters) and assuming lepton universality (5 parameters).  A summary of the combined LEP result is presented in Table~\ref{tab:lepline}.
The reader should take note of the remarkable precision of these measurements.  The large statistics and precise energy calibration provided by the resonant depolarization technique yield a measurement of the $Z$ mass to two parts in one hundred thousand!  Except for the forward-backward asymmetries (which are statistics limited at the few percent level), all of the other parameters are determined with fractional uncertainties at the $10^{-3}$ level.

\begin{table}[htb]
\begin{center}
\begin{tabular}{|l||c|} \hline
  Parameter      & Average Value \\
  \hline \hline
  $\MZ (\GeV)$ &$ 91.1871\pm0.0021$\\
  $\GZ (\GeV)$ &$  2.4944\pm0.0024$\\
  $\shad (\nb)$ & $41.544\pm0.037$\\
\hline
\hline
  $\Ree$        &$ 20.803\pm0.049 $\\
  $\Rmu$        &$ 20.786\pm0.033 $\\
  $\Rtau$       &$ 20.764\pm0.045 $\\
  $\Afbze$      &$ 0.0145\pm0.0024 $\\
  $\Afbzm$      &$ 0.0167\pm0.0013 $\\
  $\Afbzt$      &$ 0.0188\pm0.0017 $\\
\hline
\hline
 $\RZ$         &$20.768\pm0.024$\\
 $\Afbpol$     &$0.01701\pm0.00095$\\
  \hline
\end{tabular}
\caption{The combined result of the LEP lineshape analyses with (bottom box) and without (middle box) the assumption of lepton universality.}
\label{tab:lepline}
\end{center}
\end{table}

The lineshape parameter measurements are sensitive to systematic uncertainties: on the normalization of the various cross sections, on the energy scale of the machine, and on the radiative corrections that are applied to the line shape function.  The cross section normalizations are affected by uncertainties on the event selection efficiencies which are in the range $\pm$0.04-0.1\% for hadronic final states and 0.1-0.7\% for leptonic final states.  The normalizations are also sensitive to luminosity uncertainties which have experimental contributions in the range $\pm$0.033-0.09\% and theoretical contributions \cite{ref:bflward} in the range $\pm$0.054-0.06\%.  The uncertainty on the energy scale of the LEP machine leads to a $\pm$1.7~MeV uncertainty on $\MZ$ and a $\pm$1.2~MeV uncertainty on $\GZ$.  Note that the precision of the $\MZ$ determination is now limited by the energy scale uncertainty.  Uncertainties on the QED radiative corrections lead to a $\pm$0.02\% uncertainty on $\shad$ and $\pm$0.5~MeV uncertainties on $\MZ$ and $\GZ$.  

The only major change in the lineshape parameters from previous years is a $+0.9\sigma$ shift in the hadronic peak cross section.  Approximately half of this shift is directly attributable to improvements in radiative corrections.

\subsection{Tau Polarization}

\begin{figure}[htb]
\begin{center}
\epsfig{file=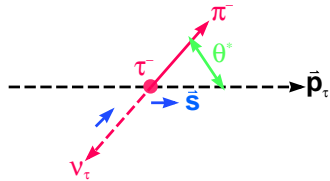,height=1.0in}
\caption{The decay of a $\tau^-$ to $\pi^-\nu_\tau$ it's rest frame.}
\label{fig:taudec}
\end{center}
\end{figure}

The LEP Collaborations determine $A_\tau$ and $A_e$ from measurements of $\polt(\cos\theta)$ as shown in equation~\ref{eq:taupol}.  The final state $\tau$-polarization is determined from the 5 decay modes: $\tau^\pm\to\pi^\pm\nu,~\rho^\pm\nu,~a_1^\pm\nu, ~e^\pm\nu\bar\nu,~\mu^\pm\nu\bar\nu$.  Since the $\tau$ decays via a pure $V-A$ current, each mode has a polarization-dependent decay distribution in laboratory variables.  For example, consider a spin polarized $\tau^-$ decaying to $\pi^-\nu_\tau$ in its rest frame as shown in Fig.~\ref{fig:taudec}.  The angular distribution of $\pi^-$ relative to the spin direction is completely asymmetric,
\begin{equation}
\frac{1}{\Gamma}\frac{d\Gamma}{d\cos\theta^*} = \left(1-\polt Q_\tau\cos\theta^*\right)/2,
\end{equation}
where $\theta^*$ is defined in the figure and $Q_\tau$ is the charge of the $\tau$.
Boosting along the spin direction (to produce a right-handed $\tau$), the distribution of scaled energy $x=E_\pi/E_b$ in the laboratory frame is given by the following expression,  
\begin{equation}
\frac{1}{\Gamma}\frac{d\Gamma}{dx} = 1-\polt Q_\tau\left(2x-1\right)=A(x)+\polt B(x),
\end{equation}
where there is a polarization-independent term $A(x)$ and linear polarization dependence with a coefficient $B(x)$.

The laboratory decay distributions of all 5 final states can be represented in the same general form,
\begin{equation}
\frac{1}{\Gamma}\frac{d\Gamma}{dx^N} =A(x_1...x_N)+\polt B(x_1...x_N),
\end{equation}
where $N=1,3,6$ for the $\ell\nu\bar\nu$, $\rho\nu$, $a_1\nu$ final states.  The final state polarization $\polt$ is extracted by fitting the decay distributions to the data.  The statistical precision obtainable with a sample of $N_{dec}$ decays for each final state can be parameterized as $\delta\polt=a_p/\sqrt{N_{dec}}$ where the analyzing powers $a_p$ and the relative number of decays for each $\tau$ final state are summarized in Table~\ref{tab:taudec}.  The total relative precision of each decay mode is shown on the bottom line.  Note that the $\pi\nu$ and $\rho\nu$ final states are the most powerful and have comparable weight.  
\begin{table}[htb]
\begin{center}
\begin{tabular}{|lccccc|}
\hline
Final State & $e\nu\bar\nu$ & $\mu\nu\bar\nu$ & $\pi\nu$ & $\rho\nu$ & $a_1\nu$ \\
\hline
\hline
Branching Ratio (\%) & 18 & 18 & 12 & 24 & 8  \\
Acceptance & 0.4 & 0.7 & 0.6 & 0.5 & 0.5  \\
Analyzing Power $a_p$ & 5 & 5 & 1.8 & 2.3 & 3.1 \\
\hline
\hline
Relative Precision & 2.7 & 2.1 & 1.0 & 1.0 & 2.2 \\
\hline
\end{tabular}
\caption{The parameters which determine the relative statistical power of each $\tau$ branching mode.  Note that the acceptances are composites of those for several experiments.}
\label{tab:taudec}
\end{center}
\end{table}

The values of $A_\tau$ and $A_e$ extracted from all five channels are summarized by experiment in Table~\ref{tab:taupol}.  The DELPHI results have been updated recently are are now final (the results from ALEPH, L3, and OPAL were already in final form).  The four-experiment averages are also given assuming that the systematic errors are uncorrelated. 
\begin{small}
\begin{table}[hbt]
\begin{center}
\begin{tabular}{|ll||c|}
\hline
Experiment & & $\cAt$ \\
\hline
\hline
ALEPH  &(90 - 95), final       & $0.1452\pm0.00052\pm0.0032\pz$  \\
DELPHI &(90 - 95), final 99    & $0.1359\pm0.0079\pm0.0055\pz$  \\
L3     &(90 - 95), final       & $0.1476\pm0.0088\pm0.0062\pz$  \\
OPAL   &(90 - 94), final       & $0.1340\pm0.0090\pm0.0100\pz$  \\
\hline
\hline
LEP Average &                    & $0.1425\pm0.0044$           \\
$\chi^2$/dof &                    & 1.3/3           \\
\hline
\hline
Experiment & & $\cAe$ \\
\hline
\hline
ALEPH   &(90 - 95), final       & $0.1505\pm0.0069\pm0.0010$  \\
DELPHI  &(90 - 94), final 99    & $0.1382\pm0.0116\pm0.0005$  \\
L3      &(90 - 95), final       & $0.1678\pm0.0127\pm0.0030$  \\
OPAL    &(90 - 94), final       & $0.1290\pm0.0140\pm0.0050$  \\
\hline
\hline
LEP Average &                     & $0.1483\pm0.0051$  \\
$\chi^2$/dof &                     & 4.8/3  \\
\hline
\end{tabular}
\caption{A summary of the coupling asymmetries $A_\tau$ and $A_e$ determined from $\tau$-polarization measurements performed by the LEP experiments.  All results are final.}
\label{tab:taupol}
\end{center}
\end{table}
\end{small}

\subsection{The Left-Right Asymmetry}
\label{sec:alr}

The SLD Collaboration at the SLAC Linear Collider has been measuring the left-right asymmetry since 1992.  It is a particularly powerful but simple and systematics-free measurement.  Because a feedback system keeps the left- and right-helicity electron currents at the SLC polarized electron source equal at the $10^{-4}$ level and because the helicity of the polarized electron beam is changed on a pulse-to-pulse basis according to a pseudorandom number sequence (which desynchronizes the the left- and right-handed currents from accelerator periodicities), the left- and right-handed luminosities are equal to excellent approximation.  This reduces the measurement to a very simple counting experiment which is described by the following equation,
\begin{equation}
A_{LR} = \frac{1}{\pole}\frac{N_Z(L)-N_Z(R)}{N_Z(L)+N_Z(R)} + \delta_{res} = \frac{1}{\pole} A_{LR}^m +\delta_{res},  \label{eq:alrmeas}
\end{equation}
where $N_Z(L)$ and $N_Z(R)$ are the number of $Z$ events (excluding final state $e^+e^-$ pairs which need a more sophisticated treatment) collected with left-handed and right-handed electron beams, respectively. In practice, only hadronic events are used in this measurement.  The leptonic final states are treated separately in a more sophisticated analysis that makes use of angular information (see section~\ref{sec:alrfb}).  The luminosity-weighted average polarization $\pole$ given in equation~\ref{eq:alrmeas} is determined from individual measurements of the
beam polarization $\poll_i$ which are associated in time with each $Z$ and used to compute the average,
\begin{equation}
\pole = (1+\xi)\frac{1}{N_Z}\sum_i^{N_Z} \poll_i
\end{equation}
where $\xi$ is a small correction (0.001-0.002) for chromatic and beam transport effects.  The term $\delta_{res}$ represents a correction for residual background and residual left-right asymmetries in luminosity, polarization, and beam energy.  The total correction is dominated by the background and left-right luminosity corrections and is at the $10^{-4}$ level.  

The operational history of the SLD experiment is summarized in Table~\ref{tab:sldhist}.  The 1997/8 run was a particularly successful one.  A total of 346,111 events were produced on the $Z$ peak (several thousand more were produced during off-peak scanning).  After 10 years of trying, the SLC instantaneous luminosity approached its design value near the end of the 1998 run.  Like the central character in a tragic opera, the accelerator had its finest day ever on 7-8 June 1998 producing over 5400 events before {\it dying} a few days prematurely due to a vacuum leak.  The entire sample consists of 557K events, most of which were logged with 0.73-0.77 beam polarization.  The SLD Collaboration released a preliminary result based upon the entire sample for this conference.
\begin{table}[htb]
\begin{center}
\begin{tabular}{|lccc|}
\hline
Year & Number of $Z$ Events & $\pole$ & $\delta\pole/\pole$ \\
\hline
1992 & 11K & 0.224$\pm$0.006 & 2.7\% \\
1993 & 50K & 0.626$\pm$0.012 & 1.7\% \\
1994/5 & 100K & 0.772$\pm$0.005 & 0.7\% \\ 
1996 & 50K & 0.765$\pm$0.004 & 0.7\% \\
1997/8 & 346K & 0.729$\pm$0.004 & 0.5\% \\ \hline
\end{tabular}
\caption{The operational history of the SLD experiment.}
\label{tab:sldhist}
\end{center}
\end{table}

The 1997/8 $\alr$ result incorporated several improvements and checks:
\begin{enumerate}
\item A number of improvements in polarimetry were realized.  Two additional independent Compton polarimeters were used to measure the backscattered photons rather than the backscattered electrons detected by the normal polarimeter.  These devices have independent calibrations and permit the reduction of the calibration uncertainty to 0.4\%.  The polarimeter was operated with interspersed high/low background running.  This permitted the comparison of polarization measurements with very different signal sizes and permitted the reduction of the uncertainty on the linearity of the system to 0.2\%.
The overall uncertainty on the polarimeter scale was reduced to $\delta\pole/\pole=0.5$\% from 0.7\%.
\item A scan of the $Z$ resonance was performed to check the energy scale of the accelerator.  The resulting $\pm$25~MeV uncertainty on the center-of-mass energy scale leads to an uncertainty on the electroweak interference corrections needed to extract $\alro$, $\delta\alr^0/\alr^0 = 0.4$\%.  This uncertainty is the second largest after the polarimeter uncertainty.
\item The polarization of the SLC positron beam was measured directly by transporting it to a M\o ller polarimeter in Endstation A of the SLAC accelerator complex.  The $e^+$ polarization was found to be consistent with zero, $\polp = -0.02\pm0.07$\%.
\end{enumerate}

The measured values of $\alr$ generated by equation~\ref{eq:alrmeas} must be corrected for electroweak interference to yield $\alro=A_e$.  The resulting values are listed in Table~\ref{tab:alrres}.  The table also lists a combined total which assumes that all systematic errors are correlated.  Note that the statistical uncertainty on the combined value is approximately twice as large as the systematic uncertainty.

\begin{table}[hbt]
\begin{center}
\begin{tabular}{|lc|}
\hline
Year & $\alro$ \\
\hline
1992 & $0.100\pm0.044\pm0.004$  \\
1993 & $0.1656\pm0.0071\pm0.0028$ \\
1994/5 & $0.15116\pm0.00421\pm0.00111$  \\
1996 & $0.15703\pm0.00573\pm0.00111$ \\ 
1997/8 & $0.14904\pm0.00240\pm0.00097$ \\ \hline
Total & $0.15108\pm0.00218$ \\ 
$\chi^2/$dof & $5.58/4$~(23\%) \\ \hline
\end{tabular}
\caption{The measured values of $\alro$ for each of the SLD runs.}
\label{tab:alrres}
\end{center}
\end{table}

The combined result is also quoted in terms of $\swein$,
\begin{eqnarray*}
\swein = 0.23101 \pm 0.00028.
\end{eqnarray*}

\subsection{Polarized Leptonic Forward-Backward Asymmetries}
\label{sec:alrfb}

The measurement of the left-right forward-backward asymmetries $\tilde A^f_{FB}$ permits the extraction of the Final State coupling asymmetries $A_f$.  For $f\neq e$, $A_f$ and $A_e$ are extracted from fits to the polar angle distributions,
\begin{equation}
\frac{d\sigma}{d\cos\theta} \propto \left(1-\pole A_e\right) \left(1+\cos^2\theta\right) + 2A_f\left(A_e-\pole\right)\cos\theta. \label{eq:polang}
\end{equation}
Final state $e^+e^-$ events must be fit to a more sophisticated expression that includes t-channel effects.

The SLD Collaboration has used this technique to measure $A_e$, $A_\mu$, and $A_\tau$ from samples of 15K~$e^+e^-$, 12K~$\mu^+\mu^-$, and 12K~$\tau^+\tau^-$ events.  The best fits to the samples 
yield the following results,
\begin{eqnarray*}
A_e &=& 0.1558\pm0.0064 \\
A_\mu &=& 0.137\pm0.016 \\
A_\tau &=& 0.142\pm0.016, 
\end{eqnarray*} 
where the much more precise determination of $A_e$ comes mostly from the first term on the right-hand side of equation~\ref{eq:polang} which effectively uses the leptonic final states to measure $\alr$ independently of the hadronic final states described in section~\ref{sec:alr}.
These measurements are consistent with lepton universality ($\chi^2/{\mathrm dof} = 1.6/2$) and can be combined into another reasonably precise measurement of $A_\ell$,
\begin{eqnarray*}
A_\ell = 0.1523\pm0.0057.
\end{eqnarray*}

\subsection{$R_b$ and $R_c$ Measurements}
\label{sec:rbrc}

The measurement of the hadronic branching ratios $R_b$ and $R_c$ is conceptually simple.  One applies some heavy quark tagging criteria to a sample of hadronic $Z$ decays, measures the fraction events which satisfy the criteria, and corrects for the efficiency of the criteria.  Unfortunately, it is difficult to determine the efficiency of a tagging procedure at the desired $<$1\% level and the more sophisticated double-tag or multi-tag approaches are used to perform the most precise measurements.  

The double-tag approach is fairly simple (the multi-tag approaches are extensions of the technique when several different tagging techniques are used for each quark flavor).  Each event in a sample of hadronic $Z$ decays is separated into two thrust hemispheres (the event is bisected by a plane normal to the thrust axis).  The $b$-quark and $c$-quark tags (which are designed to be exclusive of each other) are applied separately to each hemisphere.  The fractions of hemispheres that satisfy the $b$-tag and $c$-tag  criteria ($b_s$ and $c_s$) and the fractions of events that have double $b$-tags, $c$-tags, and mixed-tags ($b_d$, $c_d$, and $m$) are determined.  The tag-fractions are related to $R_b$ and $R_c$ by the following equations,
\begin{eqnarray} 
&&\left . \begin{array}{l}
b_s =\eff_bR_b+\eff_cR_c+\eff_{uds}\left(1-R_b-R_c\right) \\
b_d = \eff_b^dR_b+\eff_c^dR_c+\eff_{uds}^d\left(1-R_b-R_c\right) 
\end{array} \right \}
\label{eq:rb} \\
&&\left . \begin{array}{l}
c_s = \eta_bR_b+\eta_cR_c+\eta_{uds}\left(1-R_b-R_c\right) \\
c_d = \eta_b^dR_b+\eta_c^dR_c+\eta_{uds}^d\left(1-R_b-R_c\right) \\
m = 2\left[\eff_b\eta_bR_b+\eff_c\eta_cR_c+\eff_{uds}\eta_{uds}\left(1-R_b-R_c\right)\right] 
\end{array} \right \} \label{eq:rc}
\end{eqnarray}
where: the efficiencies $\eff_x$ are the probabilities that a hemisphere of type $x=b$, $c$, or $uds$ satisfy the $b$-tagging criteria; the efficiencies $\eff_x^d$ are the probabilities that events of type $x$ are double $b$-tagged; the efficiencies $\eta_x$ are the probabilities that a hemisphere of type $x$ satisfy the $c$-tagging criteria; and the efficiencies $\eta_x^d$ are the probabilities that events of type $x$ are double $c$-tagged.  In general, the double-tagging efficiencies are given by the squares of the single hemispheres but a correction for hemisphere correlations must be included,
\begin{equation}
\eff_x^d = \eff_x^2 + \lambda_x \qquad \eta_x^d = \eta_x^2 + \lambda_x^\prime
\end{equation}
where $\lambda_x$ and $\lambda_x^\prime$ account for hemisphere correlations.  Since the light-quark efficiencies $\eff_{uds}$, $\eta_{uds}$ and the {\it wrong-type} efficiencies $\eff_c$ and $\eta_b$ in most analyses are small, the corresponding correlations are omitted ($\lambda_c=\lambda_b^\prime=\lambda_{uds}=\lambda_{uds}^\prime=0$).

Most measurements of $R_b$ do not incorporate an $R_c$ analysis.  These measurements fix the value of $R_c$ to that predicted by the MSM and utilize Monte Carlo simulations to calculate $\eff_c$, $\eff_{uds}$, and $\lambda_b$.  Equations~\ref{eq:rb} are then solved for $R_b$ and $\eff_b$.  For the measurements that include $R_c$, the Monte Carlo is used to calculate $\eta_{uds}$, $\lambda_b$, and $\lambda_c^\prime$.  Equations~\ref{eq:rb} and \ref{eq:rc} are then solved for $R_b$, $\eff_b$, $\eta_b$, $R_c$, and $\eta_c$.    Note that the large efficiencies are determined from the data themselves and the corresponding systematic uncertainties are much smaller than those obtainable from the single-tag technique.

\begin{figure} [hbt]
\begin{center}
\epsfig{file=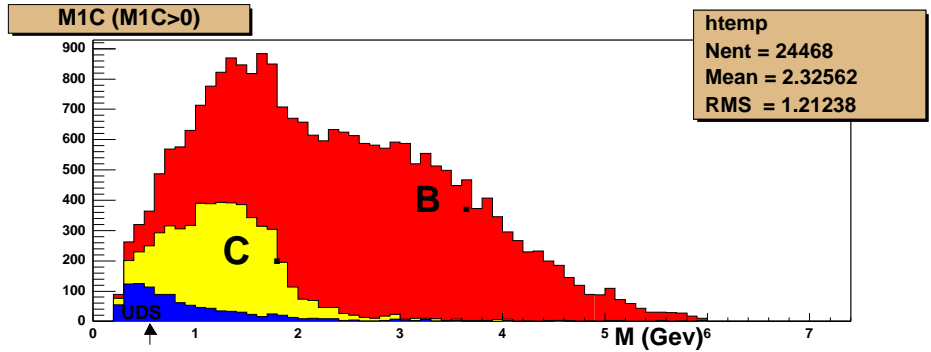,height=2.0in}
\caption{The mass distribution and quark content of a sample of topologically reconstructed vertices from the SLD.  The quark content is estimated using a Monte Carlo simulation.}
\label{fig:vtxmass}
\end{center}
\end{figure}

A number of techniques are used to tag final-state $b$- and $c$-jets.  The $b$-jet tags fall into several categories:
\begin{enumerate}
\item
Large $P$ and $P_T$ leptons from semileptonic $b$ decays,
\item
Event shape variables which are sensitive to the large mass of the $b$,
\item
``Lifetime'' tags which make use of high precision tracking systems in two different ways:
\begin{enumerate}
\item
``Traditional'' tags which require large impact parameter tracks or poor fits to a single vertex hypothesis.
\item
``Topological'' tags which are based upon reconstructed secondary or tertiary vertices.  These tags can be enhanced by requirements on vertex mass and energy.  A mass plot for events tagged by the SLD topological vertex tag is shown in Fig.~\ref{fig:vtxmass}.
\end{enumerate}
\end{enumerate}
The $c$-jet tags fall into a similar set of categories:
\begin{enumerate}
\item
Large $P$ and $P_T$ leptons from semileptonic $c$ decays are often used in conjunction with a $b$-quark analysis,
\item
Exclusively or inclusively reconstructed $D/D^*$ mesons are used to signal the presence of a primary $c$ quark,
\item
Topological lifetime tags enhanced mass and momentum requirements are very efficient.
\end{enumerate}

The current state of $R_b$ and $R_c$ measurements by the five $Z$-pole experiments is summarized in Table~\ref{tab:rbrc}.  Note that the $R_b$ measurement by DELPHI \cite{ref:delphirb} is new and is also the most precise.  The SLD result has been updated and all others are unchanged.  The $R_b$ world average, $0.21642\pm0.00073$, is approximately 0.8~$\sigma$ larger than the MSM prediction.  Both DELPHI and SLD also have updated $R_c$ results (all others are unchanged).  The $R_c$ world average, $0.1674\pm0.0038$, is about 1.3~$\sigma$ smaller than the MSM prediction.  Both quantities appear to be consistent with the MSM.

\begin{small}
\begin{table}[htb]
\begin{center}
\begin{tabular}{|cccc|}
\hline
Experiment & Technique & Sample & $R_b$ \\
\hline
ALEPH & multi-tag & 1992-1995 & $0.2159\pm0.0009\pm0.0011$  \\
DELPHI & multi-tag & 1992-1995 & $0.21634\pm0.00067\pm0.00060$ \\
L3 & multi-tag & 1994-1995 & $0.2174\pm0.0015\pm0.0028$  \\
OPAL & multi-tag & 1992-1995 & $0.2176\pm0.0011\pm0.0014$ \\ 
SLD & topol-vtx & 1993-1998 & $0.21594\pm0.00139\pm0.00140$ \\ \hline
Total & & & $0.21642\pm0.00073$ \\ 
MSM Prediction & & & $0.21579\pm0.00018$ \\ \hline \hline
Experiment & Technique & Sample & $R_c$ \\ \hline
ALEPH & lepton & 1992-1995 & $0.1675\pm0.0062\pm0.0103$  \\
ALEPH & $D^*$ Incl/Excl & 1990-1995 & $0.166\pm0.012\pm0.009$  \\
DELPHI & $D^*$ Incl/Excl & 1991-1995 & $0.161\pm0.010\pm0.009$  \\
OPAL & $D^*$ Incl/Excl & 1991-1995 & $0.180\pm0.010\pm0.012$  \\
ALEPH & $D$ Incl/Excl & 1990-1995 & $0.173\pm0.014\pm0.009$  \\
SLD & topol-vtx & 1993-1998 & $0.169\pm0.005\pm0.004$  \\
ALEPH & charm count & 1991-1995 & $0.1738\pm0.0047\pm0.0113$  \\
DELPHI & charm count & 1991-1995 & $0.1692\pm0.0047\pm0.0097$  \\
OPAL & charm count & 1991-1993 & $0.167\pm0.011\pm0.012$  \\ \hline
Total & & & $0.1674\pm0.0038$ \\ 
MSM Prediction & & & $0.17228\pm0.00006$ \\ \hline 
\end{tabular}
\caption{Summaries of the $R_b$ and $R_c$ measurements.  The MSM predictions are for $m_t=174.3\pm5.1$~GeV and $m_H=100$~GeV.}
\label{tab:rbrc}
\end{center}
\end{table}
\end{small}

\subsection{Forward-Backward Asymmetries with Quark Final States}
\label{sec:afbq}

In order to measure the angular distributions of $b\bar b$, $c\bar c$, and $s\bar s$ final states it is necessary to: tag the event flavor, measure the polar angle of the thrust axis (which is usually taken to be the $Q\bar Q$ axis), and identify which hemisphere contains the quark (as opposed to antiquark).
The tagging and $Q$/$\bar Q$ separation is achieved via several techniques which vary with the final state under study:
\begin{enumerate}
\item The tagging of $b\bar b$-events was discussed in Section~\ref{sec:rbrc}.  The separation of the $b/\bar b$ hemispheres is achieved by several techniques:
\begin{enumerate}
\item
Events which are tagged with large $P$/$P_T$ leptons are automatically sign-selected from the charge of the tagging lepton (from the decays $b\to c\ell^-\bar\nu_\ell$ and $\bar b\to\bar c\ell^+\nu_\ell$).
\item
Events which are tagged with lifetime techniques can be sign-selected by making use of the jet-charge technique.  There are several variations of the jet-charge technique.  What follows is one variant.  The forward-backward jet charge of an event is constructed by weighting the charges of each track in the forward- and backward-hemispheres by a function of the component track momentum along the thrust axis to form the quantities, $Q_F^f$ and $Q_B^f$,
\begin{equation}
Q_F^f=\frac{\sum_i^{{\vec p}_i\cdot\hat z>0} |{\vec p}_i\cdot \vec T|^\kappa q_i} {\sum_i^{{\vec p}_i\cdot\hat z>0} |{\vec p}_i\cdot \vec T|^\kappa} \qquad\qquad
Q_B^f=\frac{\sum_i^{{\vec p}_i\cdot\hat z<0} |{\vec p}_i\cdot \vec T|^\kappa q_i} {\sum_i^{{\vec p}_i\cdot\hat z<0} |{\vec p}_i\cdot \vec T|^\kappa},
\end{equation}
where the exponent $\kappa$ is usually chosen to be 1/2.  The forward-backward jet charge is then the difference of $Q_F^f$ and $Q_B^f$ and proportional to the forward-backward asymmetry $A_{FB}^f$,
\begin{eqnarray*}
Q_{FB}^f=Q_F^f-Q_B^f = \delta_f A_{FB}^f,
\end{eqnarray*}
where the hemisphere charge separation $\delta_f$ is extractable from the sums and differences of the thrust hemisphere jet-charges.  This technique is self-calibrating except for hemisphere correlation effects which must be calculated from Monte-Carlo simulations.
\item
Another technique that can be used with lifetime-tagged events is the $K^\pm$ charge sum.  If the experiment has a particle identification system, the charges of $K^\pm$ from the $b\to c\to s$ cascade can be used to separate the $b/\bar b$ hemispheres (in practice, one sums the all of the $K^\pm$ charges in each hemisphere).  As in the case of jet-charge, the requirement that the two hemispheres have consistent charges allows one to self-calibrate the analyzing power of the technique.
\item
Finally, a technique that works with topological vertex tagging is to calculate the net charge of tracks associated with a reconstructed secondary vertex.  As for jet-charge and the kaon charge sum, the requirement that the two hemispheres have consistent charges allows one to self-calibrate the analyzing power of the technique.
\end{enumerate}
\item
The tagging of $c\bar c$-events was discussed in Section~\ref{sec:rbrc}.  The separation of the $c/\bar c$ hemispheres is achieved by techniques similar to those used for $b\bar b$ sign identification:
\begin{enumerate}
\item
Events which are tagged with large $P$/$P_T$ leptons are automatically sign-selected from the charge of the tagging lepton (from the decays $c\to s\ell^+\nu_\ell$ and $\bar c\to\bar s\ell^-\bar\nu_\ell$).
\item
The sign of a $D/D^*$-tagged hemisphere can be determined from the charge of the reconstructed $D^\pm$.
\item
The net charge of tracks associated with a reconstructed secondary (topological) vertex determines the sign of the $c$-hemisphere.  As in the case of $b$-events, this technique self-calibrates.
\item
As in the case of $b$-events, charge of $K^\pm$ from the $c\to s$ cascades separates $c/\bar c$ hemispheres (the sum the $K^\pm$ charges in each hemisphere is taken).  This techniques also self-calibrates.
\end{enumerate}
\item The SLD Collaboration has recently updated a measurement of the polarized forward-backward asymmetry for $s\bar s$ final states.  The following is a summary of the event selection and hemisphere-signing technique which has a purity of 50\% to 73\% depending upon final state.  Note that this technique also self-calibrates the analyzing power.
\begin{enumerate}
\item
Events with detached vertices are removed from the sample to suppress $b\bar b$ and $c\bar c$ events.
\item
All events are required to contain fast strange particles: either $K^\pm$ with momentum larger than 9~GeV; or $K^0_s$, $\Lambda$, or $\bar\Lambda$ with momentum larger than 5~GeV
\item
A fast strange particle must be found in both hemispheres and there must be at least one signed-hemisphere ($K^\pm$, $\Lambda/\bar\Lambda$) where the largest momentum strange particle is used to sign the hemisphere.
\end{enumerate}
\end{enumerate}

\subsubsection{Unpolarized $Q\bar Q$ Asymmetries}
\label{sec:unpolafb}

The LEP Collaborations measure the $b$-quark and $c$-quark forward-backward asymmetries by fitting the angular distributions (or measuring $Q_{FB}$) of heavy quark samples.  It is necessary to correct for the charge-signing analyzing powers, backgrounds, $B^0$-${\bar B}^0$ mixing, QCD effects, and electroweak interference effects.  The resulting measurements are summarized in Table~\ref{tab:afbbc}.  The only new results for the summer of 1999, are the lepton-tag results from the ALEPH Collaboration.

Note that thes unpolarized asymmetries depend linearly on $A_e$,
\begin{equation}
A_{FB}^f = 0.75 A_e A_f, 
\end{equation}
and are therefore quite sensitive to the effective weak mixing angle (the quark asymmetries $A_Q$ are very insensitive to it). They are normally used to determine $\swein$.  In order to isolate the final state coupling asymmetries $A_b$ and $A_c$, it is necessary to divide the FB asymmetries by an average value of $A_\ell$ determined from $A_{FB}^\ell$, ${\cal P}_\tau$, and $A_{LR}$.

\begin{table}[htb]
\begin{center}
\begin{tabular}{|cccc|}
\hline
Experiment & Technique & Sample & $A_{FB}^b$ \\
\hline
ALEPH & leptons & 1991-1995 & $0.0949\pm0.0040\pm0.0023$  \\
DELPHI & leptons & 1991-1995 & $0.0998\pm0.0065\pm0.0029$ \\
L3 & leptons & 1990-1995 & $0.0960\pm0.0066\pm0.0033$  \\
OPAL & leptons & 1990-1995 & $0.0910\pm0.0044\pm0.0020$ \\ 
ALEPH & jet-chg & 1991-1995 & $0.1017\pm0.0038\pm0.0032$  \\
DELPHI & jet-chg & 1992-1995 & $0.0982\pm0.0047\pm0.0016$ \\
L3 & jet-chg & 1991-1995 & $0.0931\pm0.0101\pm0.0055$  \\
OPAL & jet-chg & 1991-1995 & $0.1004\pm0.0052\pm0.0044$ \\ \hline
Total & & & $0.0988\pm0.0020$ \\ \hline \hline
Experiment & Technique & Sample & $A_{FB}^c$ \\ \hline
ALEPH & leptons & 1991-1995 & $0.0562\pm0.0053\pm0.0036$  \\
DELPHI & leptons & 1991-1995 & $0.0770\pm0.0113\pm0.0071$ \\
L3 & leptons & 1990-1991 & $0.0784\pm0.0370\pm0.0250$  \\
OPAL & leptons & 1990-1995 & $0.0595\pm0.0059\pm0.0053$ \\ 
ALEPH & $D^*$ & 1991-1995 & $0.063\pm0.009\pm0.003$  \\
DELPHI & $D^*$ & 1992-1995 & $0.0659\pm0.0094\pm0.0035$ \\
OPAL & $D^*$ & 1990-1995 & $0.0630\pm0.0120\pm0.0055$ \\ \hline
Total & & & $0.0692\pm0.0037$ \\ \hline
\end{tabular}
\caption{Summaries of the $A_{FB}^b$ and $A_{FB}^c$ measurements.}
\label{tab:afbbc}
\end{center}
\end{table}

\subsubsection{Polarized $Q\bar Q$ Forward-Backward Asymmetries}

\begin{figure}[htb] \centering
\begin{tabular}{p{3.00in}p{3.00in}}

\begin{minipage}[htb]{3.00in} 
\begin{center}
\mbox{}
\epsfig{file=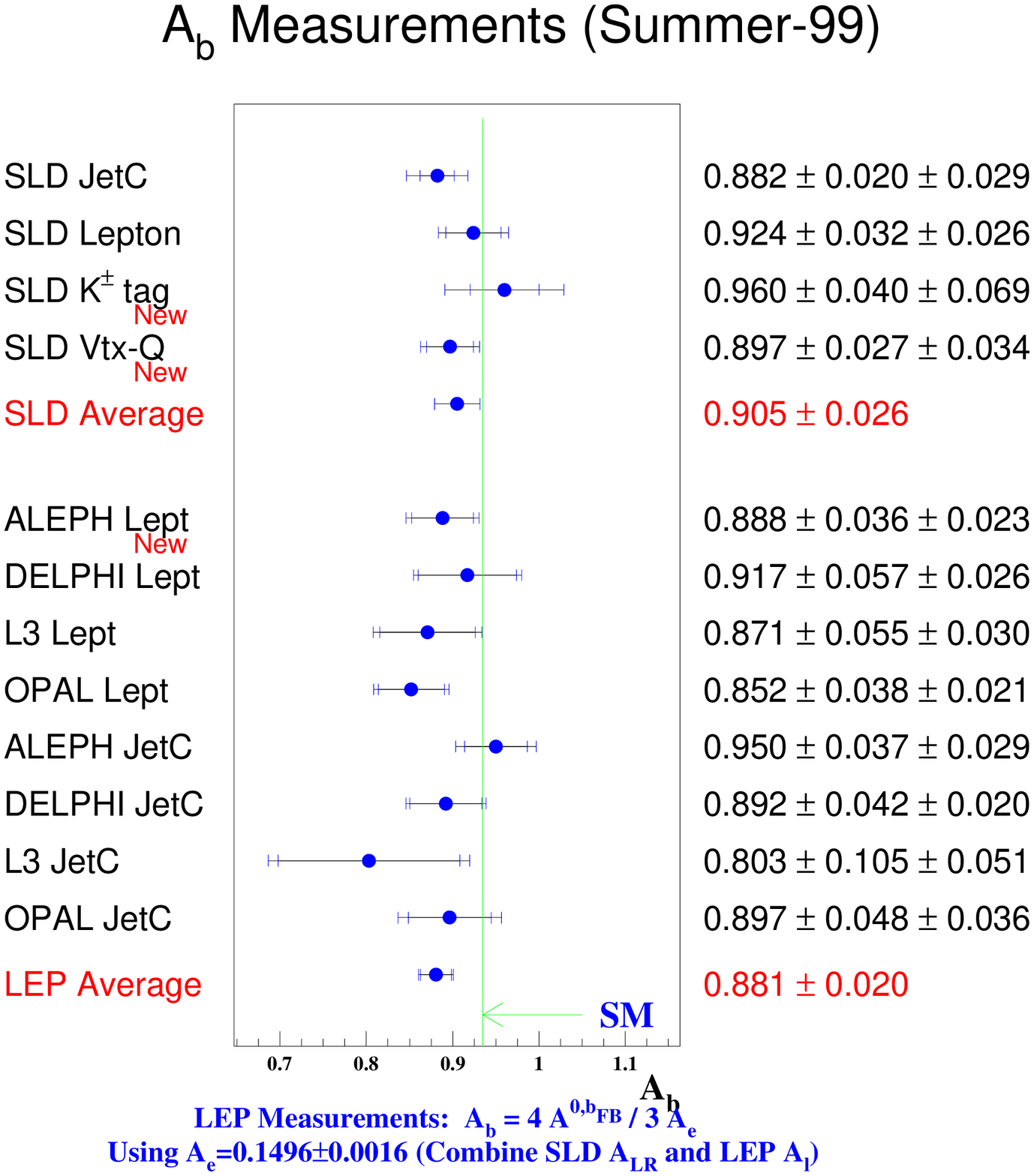,width=3.00in}
\end{center}
\end{minipage}
&
\begin{minipage}[htb]{3.00in} 
\begin{center}
\mbox{}
\epsfig{file=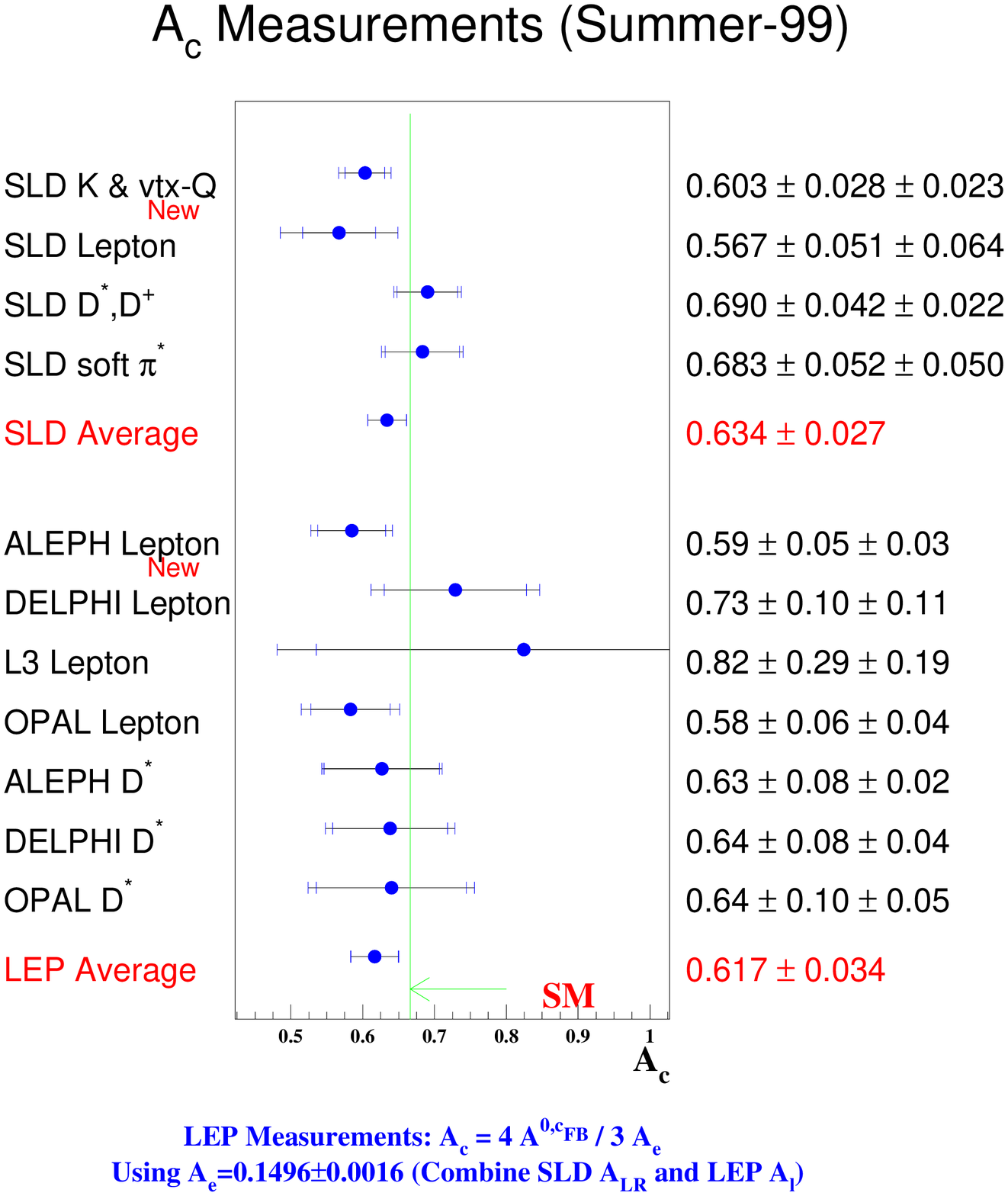,width=3.00in}
\end{center}
\end{minipage}
\\[-3ex]
\begin{minipage}[t]{2.5in}\centering \sloppy
\caption{Summary of direct and indirect determinations of $A_b$.}
\label{fig:ab}
\end{minipage}
&
\begin{minipage}[t]{2.5in}\centering 
\caption{Summary of direct and indirect determinations of $A_c$.}
\label{fig:ac}
\end{minipage}

\end{tabular}
\end{figure}

\begin{figure} [!h]
\begin{center}
\epsfig{file=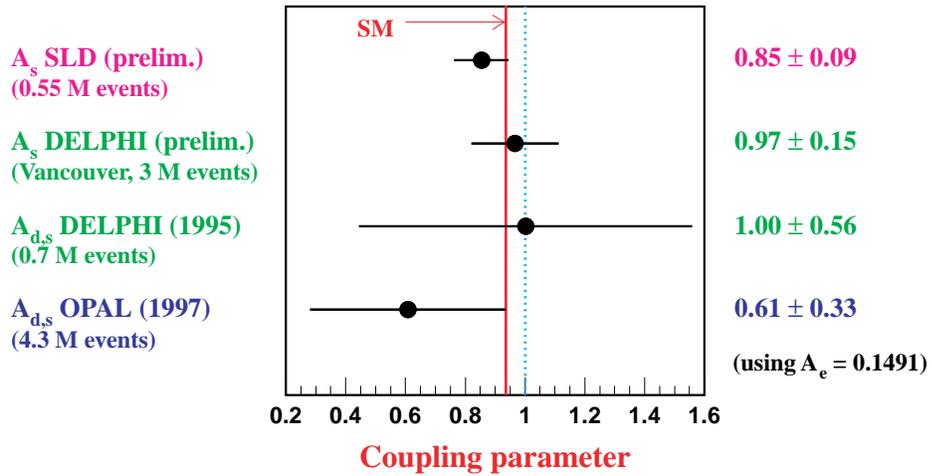,height=2.5in}
\caption{Summary of direct and indirect determinations of $A_s$.}
\label{fig:as}
\end{center}
\end{figure}

The left-right forward-backward asymmetries measured by the SLD Collaboration have been used to directly isolate the final state coupling asymmetries $A_b$, $A_c$ and $A_s$.  Using the techniques discussed in sections~\ref{sec:afbq} and \ref{sec:unpolafb}, the $A_Q$ measurements are summarized in Figs.~\ref{fig:ab}-\ref{fig:as}.  The new 1999 measurements are indicated on the figures.  The LEP $A_{FB}^Q$ measurements have been converted into $A_Q$ determinations as discussed in Section~\ref{sec:unpolafb} and are included in the figures.  Note that the direct and indirect determinations of $A_b$ are consistent with one another but (when combined) fall about 2.7~$\sigma$ below the MSM prediction (which is shown as a dotted line).  Similar behavior is observed with the $A_c$ measurements except that the discrepancy with the MSM is only about 1.9~$\sigma$.  The $A_s$ measurements have large uncertainties and are consistent with the MSM.  Note that modest inconsistency of $A_b$ with the MSM has consequences for the determination of $\swein$ which are discussed in Section~\ref{sec:swein}.
\newpage
\section{Interpretation}
\label{sec:interpretation}

\subsection{Extracted Lineshape Parameters}
\label{sec:extractedlp}

The LEP Electroweak Working Group uses the measured lineshape parameters to determine the partial decay widths of the $Z$ to leptons, hadrons, and invisible particles.  The analysis is performed with and without the assumption of lepton universality.  The results of the analysis are summarized in Table~\ref{tab:extractedlp}.  

\begin{table}[htb]
\begin{center}
\begin{tabular}{|lr||r@{$\pm$}l|}
\hline
\multicolumn{4}{|c|}{Without Lepton Universality}        \\
\hline
\hline
 $\Gee    $ & $ (\MeV)$ & $83.90$ & $0.12$ \\
 $\Gmumu  $ & $ (\MeV)$ & $83.96$ & $0.18$ \\
 $\Gtautau$ & $ (\MeV)$ & $84.05$ & $0.22$ \\
\hline
\hline
\multicolumn{4}{|c|}{With Lepton Universality}        \\
\hline
\hline
 $\Gll $ & $ (\MeV) $ & $   83.96$ & $0.09$ \\
 $\Ghad$ & $ (\MeV) $ & $ 1743.9 $ & $2.0 $ \\
 $\Ginv$ & $ (\MeV) $ & $  498.8 $ & $1.5 $ \\
\hline
\end{tabular}
\caption{The extracted lineshape parameters as determined by the LEP Electroweak Working Group.}
\label{tab:extractedlp}
\end{center}
\end{table}

We note that the leptonic widths are consistent with lepton universality.  The invisible width is of particular interest since it is sensitive to the presence of ``invisible'' final states.  The ratio of the invisible and leptonic widths is found to be
\begin{eqnarray*}
\Ginv/\Gll = 5.941 \pm 0.016.
\end{eqnarray*}
Dividing this number by the MSM value for the ratio of neutrino and leptonic widths $\Gamma_{\nu\nu}/\Gll=1.9912\pm0.0012$ yields the number of light neutrinos,
\begin{eqnarray*}
N_\nu = 2.9835\pm0.0083.
\end{eqnarray*}
Note that $N_\nu$ is about 2~$\sigma$ less than the expected value of 3.  In previous determinations it had been consistent with $N_\nu=3$.  The shift is largely a consequence of the shift in the peak hadronic cross section discussed in Section~\ref{sec:zlineshape} which was due, at least in part, to improvements in radiative corrections.
Assuming that the number of light neutrinos is 3, the invisible width is converted into a 95\% upper limit on additional invisible width $\Delta\Ginv$,
\begin{eqnarray*}
\Delta\Ginv<2.0~{\rm MeV}.
\end{eqnarray*}

\subsection{Tests of Lepton Universality}
\label{sec:leptonuniv}

\begin{figure} [!h]
\begin{center}
\epsfig{file=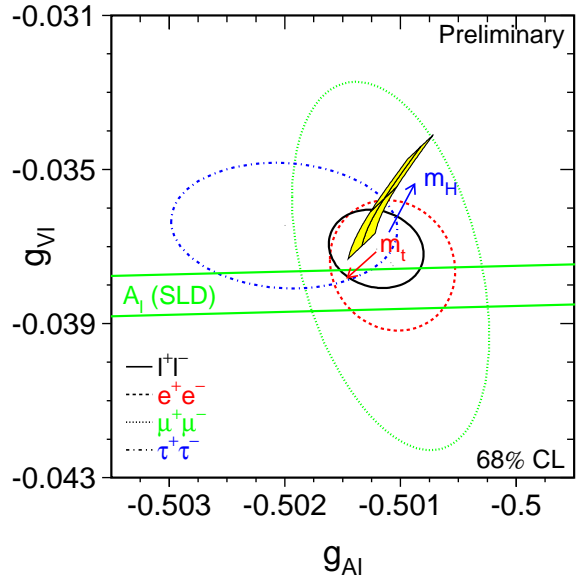,height=3.0in}
\caption{The 68\% confidence regions in $g_{V\ell}$-$g_{A\ell}$ for each lepton flavor.  The solid curve is for all flavors combined and the MSM-allowed region is also shown.}
\label{fig:leptuniv}
\end{center}
\end{figure}
The various measurements of $\Gll$, $A_{FB}^\ell$, and $A_\ell$ have been used by the LEP Electroweak Working Group to unfold values of $g_{V\ell}=v_\ell/2$ and $g_{A\ell}=a_\ell/2$ for each lepton flavor.  The resulting determinations of $g_{V\ell}$ and $g_{A\ell}$ are shown for each flavor in Fig.~\ref{fig:leptuniv}.  It is clear that the couplings of the different flavors are consistent with each other and with the MSM which is shown as the banana-shaped region.  The analysis includes most of the information reported in this talk except for the SLD measurements of the leptonic left-right forward-backward asymmetries.  Since the SLD measurement of $A_\mu$ is the best in the world, its inclusion should noticeably improve the determination of $g_{V\mu}$.  The results can also be presented in the form of ratios of the vector and axial-vector couplings,
\begin{eqnarray*}
\frac{g_A^\mu}{g_A^e}&=&1.0001\pm 0.0014 \ \ \ \ \ \ \frac{g_V^\mu}{g_V^e} = 0.981\pm0.082 \\
\frac{g_A^\tau}{g_A^e}&=&1.0019\pm 0.0015 \ \ \ \ \ \ \frac{g_V^\tau}{g_V^e} = 0.964\pm0.032,
\end{eqnarray*}
which are all consistent with universality.  Note that the precisely measured partial widths constrain the ratios of the (large) axial-vector couplings at the $10^{-3}$ level whereas the determinations of the coupling asymmetries constrain the ratios of the (small) vector couplings at the few percent level.

\subsection{The Effective Weak Mixing Angle and $A_b$}
\label{sec:swein}

\begin{figure} [hbt]
\begin{center}
\epsfig{file=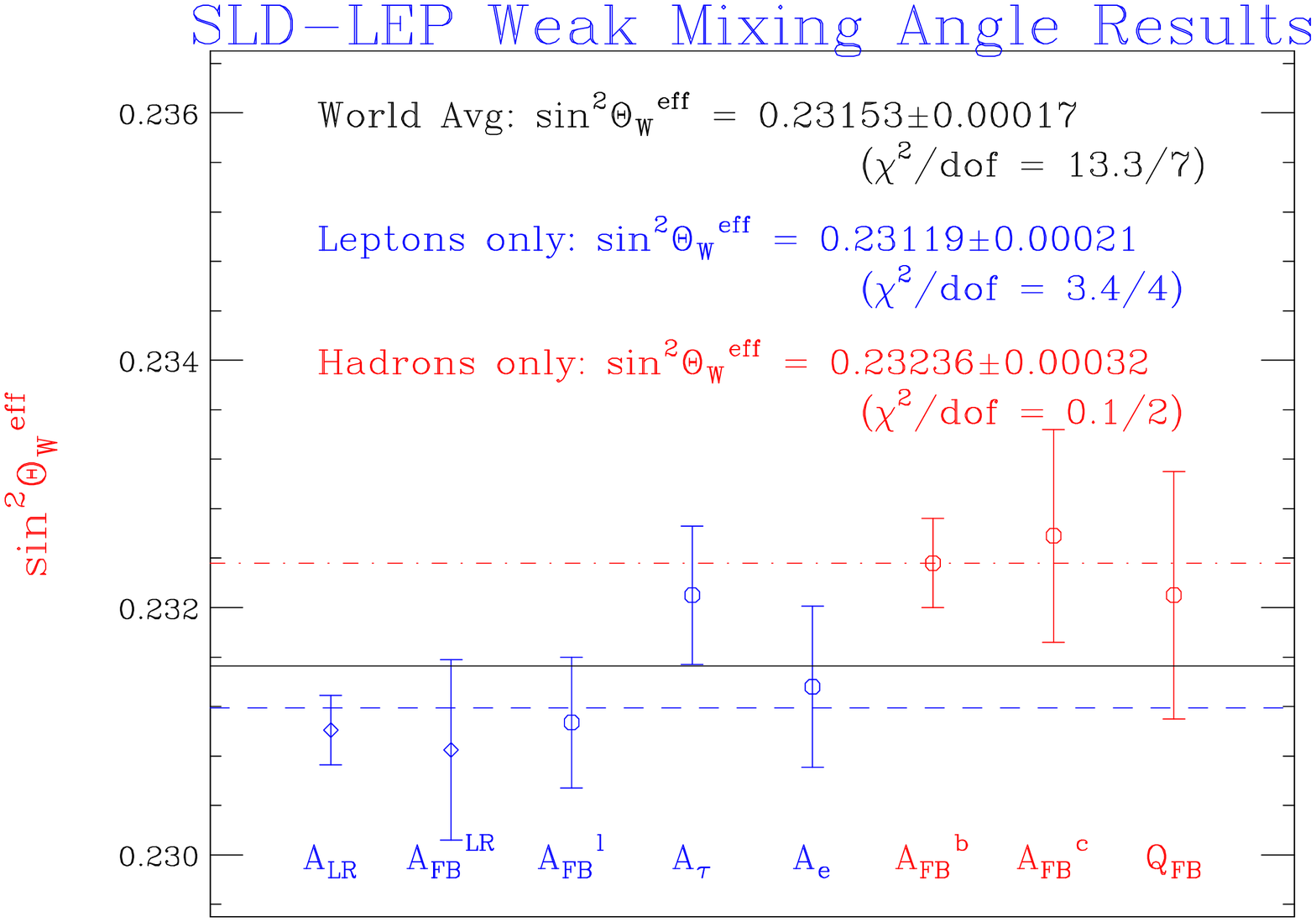,height=3.0in}
\caption{A summary of all of the determinations of $\swein$.}
\label{fig:swein}
\end{center}
\end{figure}

The $Z$-pole asymmetries that determine $A_\ell$ can equivalently be cast as determinations of $\swein$.  A summary of all measurements of $\swein$ is presented in Fig.~\ref{fig:swein}.  The purely leptonic determinations (which do not depend upon the quark coupling asymmetries $A_Q$) are shown on the left-hand side of the plot: $A_{LR}$, $A_\ell$ (from the leptonic left-right forward-backward asymmetry analysis), $A_{FB}^\ell$, $A_\tau$ (from $\polt$), and $A_e$ (from $\polt$).  The hadronic determinations, $A_{FB}^b$, $A_{FB}^c$, and $Q_{FB}$ (from the $Q_{FB}^f$ technique discussed in Section~\ref{sec:afbq} applied to the entire hadronic sample), are shown on the right-hand side of the plot.  The solid horizontal line shows the average of all techniques,
\begin{eqnarray*}
\swein = 0.23153\pm0.00017.
\end{eqnarray*} 
Note that the fit of the eight measurements to a single value has a $\chi^2$ per degree of freedom of 13.3/7 which has a probability of 6.5\%.  This is acceptable but may also indicate that the single-value hypothesis is flawed.  If the leptonic and hadronic determinations are averaged separately (see the dashed and dash-dotted lines in Fig.~\ref{fig:swein}), the fit probabilities improve markedly to 49\% and 95\%, respectively.  Another strange feature is that the two most precise determinations of $\swein$ are $A_{LR}$ and $A_{FB}^b$ which disagree with one another by 3.0~$\sigma$.

There are several possible explanations for this modest discrepancy:
\begin{enumerate}
\item
The deviations are simply the result of statistical fluctuations.
\item
Unknown systematic effects are distorting some or all of the results.
\item
We are seeing evidence of new physics.  Another manifestation of this discrepancy was the deviation of $A_b$ from the MSM when the world average $A\ell$ was used to extract $A_b$ from $A_{FB}^b$.  The SLD direct determination of $A_b$ is a check on this idea.  A graphical representation of this problem is shown in Fig.~\ref{fig:alab}.  The measurements of $A_\ell$, $A_b$ and $A_{FB}^b=0.75A_\ell A_b$ are plotted as 68\% confidence regions in $A_\ell$-$A_b$ space.  We note that they do tend to overlap in a region away from the MSM prediction which appears as a horizontal line (for $\Mt=174.3\pm5.1$~GeV \cite{ref:tmass} and 100~GeV$<\MH<1$~TeV).  A fit to $A_b$ and $A_\ell$ is shown as 68\% and 95\% ellipses.  The best fit for $A_b$ deviates by 2.7~$\sigma$ from the MSM.  The possibility that new physics might be affecting the $Zb\bar b$ couplings has been studied recently by M.~Chanowitz \cite{ref:chanowitz}.  He finds that it is possible to accommodate all measurements by changing both the left- and right-handed $b$ neutral current couplings.  However, there may be observable consequences in future measurements of flavor-changing neutral currents and rare $K$ decays. 
\begin{figure} [hbt]
\begin{center}
\epsfig{file=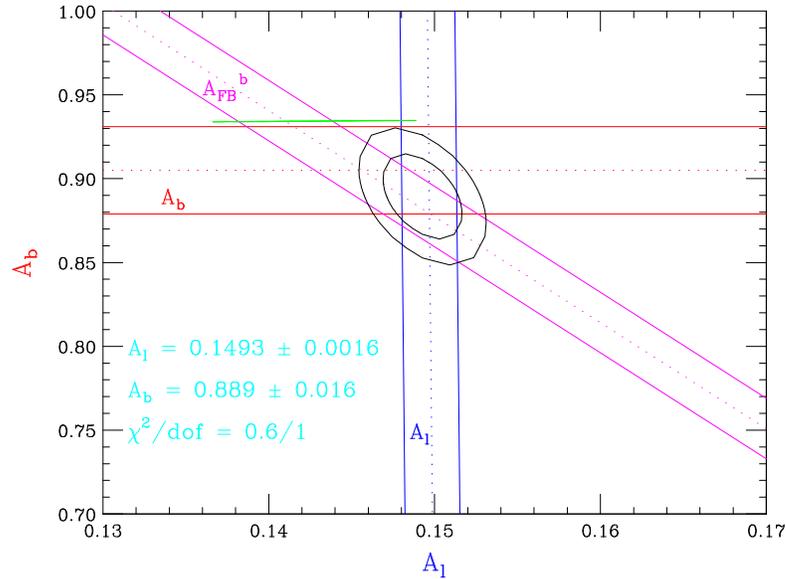,height=3.0in}
\caption{Measurements of $A_\ell$, $A_b$, and $A_{FB}^b$ are plotted in $A_\ell$-$A_b$ space.  The MSM prediction appears as a horizontal line and the best fit to the data is shown as concentric 68\% and 95\% ellipses.}
\label{fig:alab}
\end{center}
\end{figure}
\end{enumerate}

\subsection{Consistency with the Minimal Standard Model}

\begin{figure} [hbt]
\begin{center}
\epsfig{file=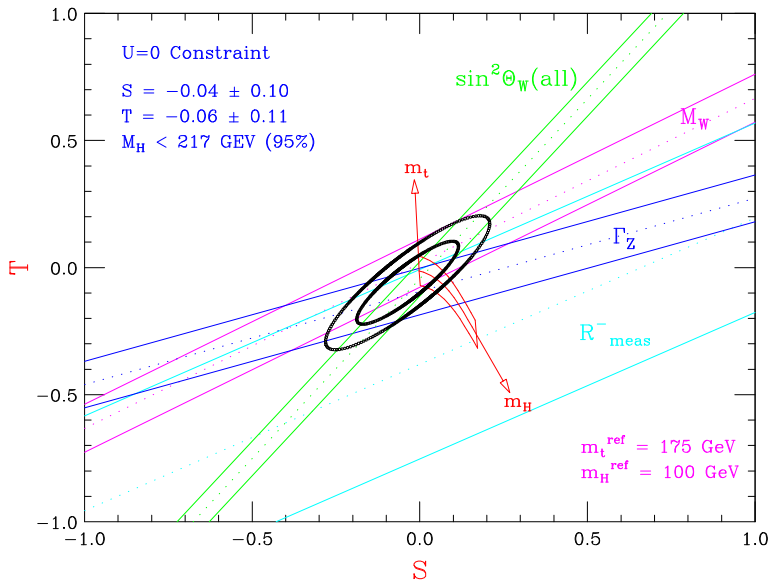,height=3.0in}
\caption{$S$-$T$ analysis of fourteen precisely measured electroweak observables.}
\label{fig:st}
\end{center}
\end{figure}

We can compare the $Z$-pole electroweak measurements with others and the Standard Model using the ``Model-Independent'' $S$ and $T$ parameters of Peskin and Takeuchi \cite{ref:pestak}.  The $S$ and $T$ parameters describe the effects of vacuum polarization corrections and can describe a more general theory so long as vertex corrections and other non-universal effects are small.  To perform the analysis, fourteen EW measurements (the $W$-mass $\MW$ \cite{ref:lepew}, the ratio $R_\nu^-$ measured in neutrino scattering \cite{ref:rnu}, the weak charges of Cesium \cite{ref:cs} and Thallium \cite{ref:th}, $\GZ$, $\shad$, $R_\ell$, $A_{FB}^\ell$, $A_\tau$, $A_e$, $\alr$, $A_{FB}^b$, $A_{FB}^c$, $Q_{FB}$) are fit to $S$, $T$, $\alpha_s$, and the hadronic part of the electromagnetic vacuum polarization correction $\Delta\alpha_{had}^5(\MZ^2$) (which is constrained to $277.5\pm1.7\times 10^{-4}$ as suggested by Kuhn and Steinhauser \cite{ref:ks}).    The results of the analysis are presented in Fig.~\ref{fig:st}.  The 68\% confidence regions for the most precisely measured observables appear as bands.  Note that all of the quantities that determine $\swein$ have been combined into a single band.  The results of the fit are shown as concentric 68\% and 95\% elliptical confidence regions.  The prediction of the MSM appears as the banana-shaped region for $\Mt=174.3\pm5.1$~GeV and 100~GeV$<\MH<1$~TeV. 

Examining Fig.~\ref{fig:st}, we can see that the $\swein$-band is the smallest and provides the most information.  Due to recent improvements, the measurement of $\MW$ now provides the second most precise input, and is slightly ahead of $\GZ$.  The data are clearly consistent with the MSM and favor a ``light'' Higgs boson.  This observation can be made more quantitative by applying $\Mt=174.3\pm5.1$~GeV as a constraint, and fitting the $S$-$T$ likelihood function for $\MH$.  The 95\% one-sided confidence limit on the Higgs mass is 217~GeV.

\begin{figure} [hbt]
\begin{center}
\epsfig{file=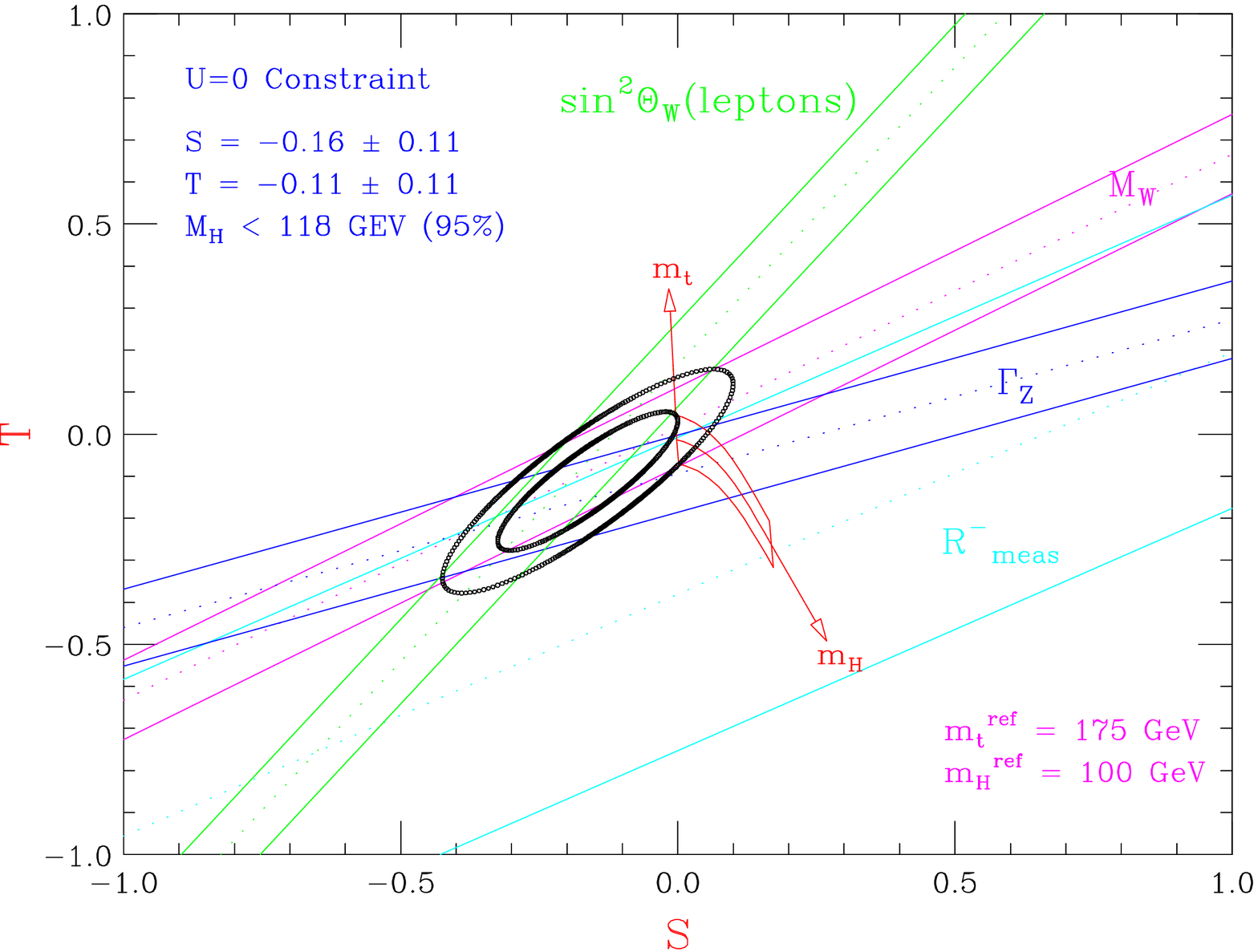,height=3.0in}
\caption{$S$-$T$ analysis of eleven precisely measured electroweak observables.}
\label{fig:stleptons}
\end{center}
\end{figure}

It is interesting to note that the possible $A_b$ anomaly discussed in Section~\ref{sec:swein} does affect these conclusions somewhat.  Excluding the three hadronic determinations of $\swein$ from the $S$-$T$ analysis, changes the picture to the one presented in Fig.~\ref{fig:stleptons}.  Note that the $S$-$T$ region determined by the leptonic measurements of $\swein$ is still the smallest, but the agreement with the MSM becomes marginal.  The Standard Model region is nearly excluded by a 95\% one-sided confidence interval in $S$.  The central value of the Higgs mass determination is 42~GeV and the one-sided 95\% upper limit is 118~GeV.

\section{Conclusions and Summary}
\label{sec:summary}

\begin{itemize}
\item The era of $Z$-pole electroweak physics has come to an end.
\item The decade has been a ``Golden Age'' for precise electroweak measurements:
\begin{itemize}
\item The $Z$ mass $\MZ$ is measured to $2.2\times10^{-5}$! It is the third most precisely determined electroweak parameter and used as an input to the theory.
\item The other lineshape parameters are measured at the $10^{-3}$ level which is an experimental tour-de-force and leads to the remarkably precise determination of the invisible width and the conclusion that there are three light neutrinos and essentially no width for other unseen processes.
\item The effective weak mixing angle $\swein$ determines most of what we know about loop-level processes.
\end{itemize}
\item There is generally good agreement with the minimal Standard Model but there are still some lingering inconsistencies with leptonic and hadronic determinations of $\swein$:
\begin{itemize}
\item The measurements of $A_b$ are anomalous by $2.7\sigma$.
\item The inconsistency has consequence for the interpretation of the data in terms of the MSM.  If there is an anomaly in the $Zb\bar b$ couplings, then the agreement of the data and MSM is marginal.
\end{itemize}
\item We are leaving the $Z$ a bit too soon.  Real improvements could be realized with more running of the SLC/SLD program. 
\item We have made enormous progress in the last 10 years.  An $S$-$T$ plot describing the state of electroweak physics 10 years ago is shown in Fig.~\ref{fig:stold}.  The current plot (Fig.~\ref{fig:st}) is shown as the dashed inset.
\begin{figure} [hbt]
\begin{center}
\epsfig{file=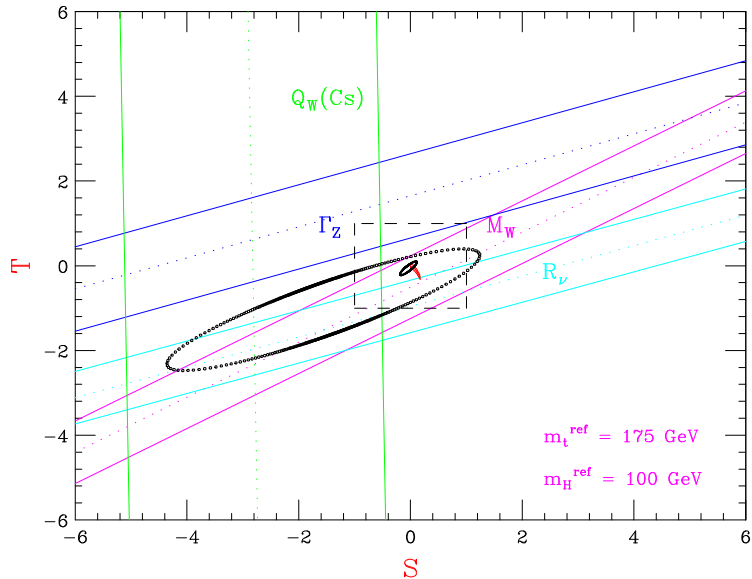,height=3.0in}
\caption{$S$-$T$ analysis describing electroweak data available at LP89.}
\label{fig:stold}
\end{center}
\end{figure}
\end{itemize}

\bigskip
The author would like to thank Bob Clare and his colleagues in the LEP Electroweak Working Group for their (usual) excellent summary of LEP results.  The author would also like to thank his colleagues in the SLD Electroweak and Heavy Flavors Working Groups.


\begin{thebibliography}{99}

\bibitem{ref:markii}
G.J.~Feldman, SLAC-PUB-5143, November 1989; Proceedings of the 14$^{th}$ International Symposium on Lepton and Photon Interactions, Stanford, CA Aug 7-12, 1989, Ed. M.~Riordan, World Scientific, p.225.

\bibitem{ref:lepew}
Most of the LEP results given in this talk are due to the LEP Electroweak Working Group, private communication.

\bibitem{Barate:1999ce}
R.~Barate {\it et al.}
[ALEPH Collaboration],
CERN-EP-99-104, July 1999.

\bibitem{ref:bflward}
B.~Ward \etal, Phys. Lett. {\bf B450}, 262 (1999).

\bibitem{ref:delphirb}
P.~Abreu \etal [DELPHI Collaboration], E. Phys. J. {\bf C10}, 415. 

\bibitem{ref:tmass}  F.~Abe \etal, Phys. Rev. Lett. {\bf 82}, 271 (1999); D.~Abbot \etal, Phys. Rev. {\bf D58}, 052001 (1998).

\bibitem{ref:chanowitz}
M.~Chanowitz, LBNL-43248, hep-ph/9905478, May 1999.

\bibitem{ref:pestak}
M.E.~Peskin and T.~Takeuchi, Phys. Rev.
Lett. {\bf 65}, 964 (1990); Phys. Rev. D {\bf 46}, 381
(1992).

\bibitem{ref:rnu} K.S.~McFarland et al., hep-ex 9806013, June 1998.

\bibitem{ref:cs} S.C.~Bennett and C.E.~Wieman, {\it Phys. Rev. Lett.} {\bf 82}, 2484 (1999).

\bibitem{ref:th} P.A.~Vetter, et al., {\it Phys. Rev. Lett.} {\bf 74}, 2658 (1995); N.H.~Edwards, et al., {\it Phys. Rev. Lett.} {\bf 74}, 2654 (1995).

\bibitem{ref:ks}
J.H.~Kuhn and M.~Steinhauser, Phys. Lett. {\bf B437}, 425 (1998).


\end{thebibliography}
\end{document}